\documentclass{aanew0}
\input psfig.tex
\usepackage{graphicx}

\newcommand{\ms}{M$_{\odot}$}
\newcommand{\zs}{Z$_{\odot}$}
\newcommand{\ddp}{D/D$_P$}
\newcommand{\dph}{D$_P$/H}

\begin{document}
   \title{Deuterium at high redshift: primordial or evolved?}

    \author{Nikos Prantzos \inst{1}, Yuhri Ishimaru \inst{1,2}}

   \offprints{N. Prantzos}

   \institute{Institut d'Astrophysique de Paris,
              98bis Bd Arago, 75014 Paris\\
         \and
             Department of Astronomy, University of Tokyo, 7-3-1 Hongo,
             Bunkyo-ku, Tokyo 113-0033 \\
             }

   \date{Received May 15, 200 ; accepted June 19, 2001 }

   \abstract{
On the basis of arguments from galactic chemical evolution we suggest that
the recent observations of D/H vs. metallicity
in several high redshift absorbers are best understood if the primordial D
value
is in the range D$_P$/H$\sim$2-3 10$^{-5}$.
This range  points to a rather high 
baryonic density ($\Omega_Bh^2$=0.019-0.026)
compatible  to the one obtained by recent estimates based on the 
Cosmic Microwave Background (CMB) anisotropy measurements. Slightly higher values
(D/H$\sim$4 10$^{-5}$) are found in Lyman limit systems. Such values
are still compatible with CMB estimates but, if taken at face value,
suggest a trend of decreasing D abundance with metallicity. We argue that
special
assumptions, like differential enrichment, are required to explain the data
in that case. A clear test of such a differential enrichment would be
an excess of products of low mass stars like C and/or N in those systems,
but currently available data
of N/Si in DLAs do not favour such a ``non-standard'' senario.
   }
 \authorrunning{N. Prantzos and Y. Ishimaru}
 \titlerunning{Deuterium at high redshift}
   \maketitle
%

\section{Introduction}

The well known sensitivity of the primordial D abundance on the
baryon-to-photon
ratio makes it the best ``baryometer'' of the Universe, as proposed in
Reeves et al. (1973). Adams (1976) suggested that QSO absorption line
systems
at high redshift offer the best opportunity to determine the primordial D
value (\dph), without having to account for its subsequent depletion by
successive
generations of stars (astration). In the past few years, observations of D
in
Lyman limit systems (LLS, with H column densities $>$3 10$^{17}$ cm$^{-2}$)
gave rather conflicting results, pointing either to very high ($>$10$^{-4}$)
or low
($<$4 10$^{-5}$) values (e.g. Lemoine et al. 2001 for a review); however,
the best determined values  favour a low \dph$\sim$3
10$^{-5}$ (e.g Burles \& Tytler 1998).

Most recently, D has been detected in Damped Lyman alpha systems (DLAs) with
higher
column densities, N(HI)$>$2 10$^{20}$ cm$^{-2}$ ( O'Meara et al. 2000,
D'Odorico et al. 2001, Pettini \& Bowen 2001). If all the data of high
redshift
systems (both LLS and DLAs) are taken into account, a trend of decreasing D
with metallicity seems to emerge (e.g. O'Meara et al. 2001). However,
Pettini \& Bowen (2001) suggest that the LLS abundance determinations may
suffer from systematic uncertainties and only the DLA values should be
considered
as reliable, which point to a quite low \dph=2.2$\pm$0.2 10$^{-5}$.

In this paper we analyse the recent data of D vs. metallicity in high
redshift
systems from the point of view of galactic chemical evolution, based
on a simple relation existing 
between D abundance and metallicity. We conclude that
the most straightforward interpretation of the data is by assuming a low
\dph=2-3 10$^{-5}$. Slightly higher values can be accounted for by making
special
assumptions, like ``differential'' galactic winds or a skewed stellar
initial
mass function (IMF),
which generically produce high abundance ratios of Carbon and Nitrogen
(typical
products of intermediate mass stars) with respect
to alpha elements (typical products of massive stars), as originally
suggested by Jedamzik \& Fuller (1997);
the fact that such values have not been
observed up to now in high redshift absorbers favors again the
``conservative''
interpretation of low primordial D/H, in agreement with the most recent
estimates from Standard Big Bang Nucleosynthesis calculations
(e.g. Burles et al. 2000).

\section {Models of deuterium vs. metal evolution}

The relationship between D abundance and metallicity $Z$ (expressed by
the abundance of a product of massive stars, like O or Si) is investigated
in
Prantzos (1996, hereafter P96). This relationship can be easily obtained
analytically in the framework of IRA (Instantaneous Recycling Approximation)
for closed models or models with outflow rate proportional to the star
formation rate (SFR). As shown in P96 (his Eq. 6), the D vs. metallicity
relation in these cases is given by:

\begin{equation}
{{D}\over{D_P}} \ = \ e^{-{{Z}\over{y}}{{R}\over{1-R}}}
\end{equation}

\noindent where $y$ is the metal yield (in the same units as $Z$)
and $R$ the return mass fraction (see e.g.
Tinsley 1980 for definitions of these quantities).

From Eq. (1) it is obvious that the D/D$_P$ vs. $Z$ relation is unique and
independent
of the history of the system, i.e it does not depend explicitly either on time,
the star formation rate or the  gas fraction $\sigma_{GAS}$,
which do not appear in the equation. It only depends on $y$ and $R$,
both of which depend on the adopted IMF; once the IMF is fixed, there are no
more degrees of freedom.

\begin{figure}
\psfig{file=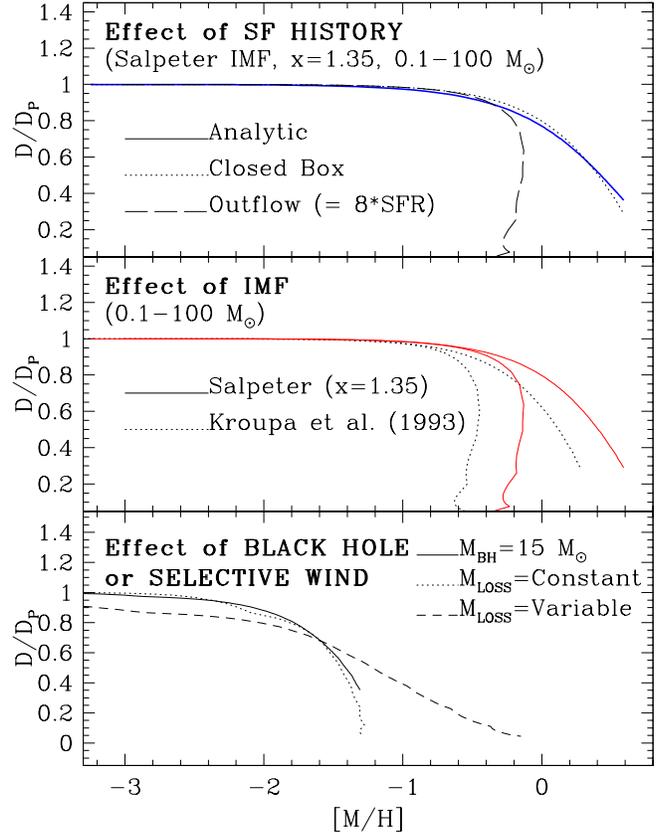,height=12.cm,width=0.5\textwidth}
\caption{\label{}
Deuterium vs. metallicity relationship, affected by various model
ingredients.
{\it Top:} Effect of the Star formation history:
an analytical Closed Box model ({\it solid line}),
a numerical Closed Box model ({\it dotted line}),
and an Outflow model ({\it dashed line}).
{\it Middle:} Effect of the IMF:
Models with the Salpeter IMF with x=1.35 ({\it solid line})
and with the Kroupa et al. (1993) IMF ({\it dotted line}).
The two curves for each IMF correspond to Closed Box
and Outflow models, respectively, as in the top panel.
{\it Bottom:} Effect of ``differential'' enrichment:
Models are based on the assumptions that stars with $M>15M_\odot$ eject no metals
({\it solid line}), or that only 1\% of the massive star ejecta is retained
in the system ({\it dotted line}), or that the retained fraction
of the massive star ejecta varies with time $t$ as $t^3$.}
\end{figure}

However, it is well known that the IRA solutions fail at very low gas
fractions
(typically, below 10 \%) and for the products of low-mass long-lived stars,
which
return their ejecta in the ISM with a long time delay (see Prantzos and
Aubert 1995
for an illustration of the effect).
In Fig. 1 (top panel) we plot the relation between D and metallicity for
several
chemical evolution models: (1) Closed Box with IRA; (2) Closed Box without
IRA
and (3) Outflow at a rate $ f =$ 8 SFR. In all cases we adopt a SFR $\psi$ =
0.25
$\sigma_{GAS}$ Gyr$^{-1}$ and the Salpeter IMF with slope x=1.35 between 0.1
and
100 \ms; this IMF leads to a return fraction $R\sim$0.31 and a yield
$y\sim$0.6
\zs \ for oxygen when the yields of Woosley \& Weaver (1995) are adopted.
It can be seen that the analytical solution matches quite well the numerical
one
for the closed box model. In the case of outflow the final gas fraction is
much
lower and the divergence from the analytical solution more important:
metallicity
reaches lower values and D depletion is more important than in the closed
box case.
Both effects are due to the impact of deuterium-free and metal-free material
returned
by low-mass stars at late times in the system, when star formation and metal
production have dropped to negligible levels.

The effect of the IMF on the D vs. metallicity relationship is illustrated
in the middle
panel of Fig. 1. The results of the numerical closed box models are
presented for a Salpeter IMF
with slope x=1.35 (same as in the top panel) and for a Kroupa et al. (1993)
IMF.
The latter has a steeper
slope in the high mass range (x=1.7 for M$>$1 \ms)
and a swallower one in the low mass range (x=1.2 for M$<$1 \ms); as a
result,
it leads to lower metal yield than the Salpeter IMF.
For the same gas fraction,
lower metallicities are obtained with the Kroupa IMF while D depletion
starts
at slightly lower metallicity than with the Salpeter IMF.

Despite some differences, the common feature of all the ``standard'' models
presented in the top and middle panels of Fig. 1 (which assume that
the ISM is polluted by the ejecta of both low and high mass stars)
is that negligible D depletion is obtained below
a metallicity Z=0.1 \zs.
This is due to the fact that for all ``standard'' IMFs one has $y\sim$0.5
\zs \
and $R\sim$0.3, i.e.
the factor $R/[y (1-R)]$ in Eq. (1) is $R/[ y (1-R)] \leq$1; for
metallicities $Z<$0.1 \zs \
the exponent in Eq. (1) is dominated by  the (very small)
factor $Z$, leading to D/D$_P\sim$1.

Obviously,  ``non-standard'' assumptions
are required in order to obtain D depletion at low metallicities, comparable
to
those observed in high redshift systems. Some possibilities are explored
in the lower panel of Fig. 1, where the duration of the evolution of our
model systems is fixed to 2 Gyr
(corresponding, roughly, to the ages of the high redshift clouds where D is
observed).
The common feature of all these models is that they assume ``differential''
enrichment
of the ISM, preferentially by low mass stars, while the metal-rich ejecta
of massive stars are lost.

(i) In the first case ({\it solid curve}), it is assumed that stars with
M$>$15 \ms \
form black holes and eject no metals; as a result, the yield $y$ is
considerably reduced,
metallicity never increases above 0.1 \zs \  and some D depletion is
obtained
even below [M/H]=-1.

(ii) In the second case ({\it dotted curve}) it is assumed that a constant
fraction $g$=0.01
of the ejecta of massive stars is retained by the system, while the rest
(0.99)
is lost from the system through  the form of
a ``selective'' galactic wind (which affects only the ejecta of supernovae
but not
of intermediate mass stars); the evolution of D vs. O is quite similar to
the one
of case (i) for obvious reasons.

(iii) In the last  case ({\it dashed curve}) it is assumed that the fraction
of the
ejecta retained by  the system  varies with time $t$ as
 $g \propto \ t^{3}$, i.e. it is quite small early on and becomes important
at late times (for instance,
because the system becomes sufficiently massive to retain the ejecta
at late times).

The parameters (M$_{BH}, g$) of the ``non-standard'' models of
cases (i) to (iii) are not based on physical arguments and
the only motivation of the underlying
assumptions is to show that substantial D depletion can take place even at
low metallicities, at least in principle.
Note that galactic winds are thought to be a crucial ingredient in models of
early-type galaxies (e.g. Larson 1974, Arimoto \& Yoshi 1987)
and they  obviously contribute to the enrichment of
intergalactic gas in galaxy clusters
(e.g. Renzini et al. 1993);
also, other systems
like blue compact dwarf galaxies may suffer such selective winds (e.g. Kunth
\& Ostlin 2000 for a review).
However, the important point is that all these assumptions have generic
observational consequences that concern not only D but other elements
as well, and which  will be discussed in Sect. 3.

At this point, we note that the low column densities
of LLSs imply that
these systems  have not have formed stars and metals themselves, 
but they have been contaminated
by nearby star forming regions. In that case, the D vs. $Z$ curves in Fig. 1
constitute lower limits to the absorber abundances of D at a given
metallicity,
the true value depending
on the degree of mixing between the two systems. Obviously, in that case it
would be even more difficult to obtain substantial D depletion at low
metallicities.

\begin{figure}
\psfig{file=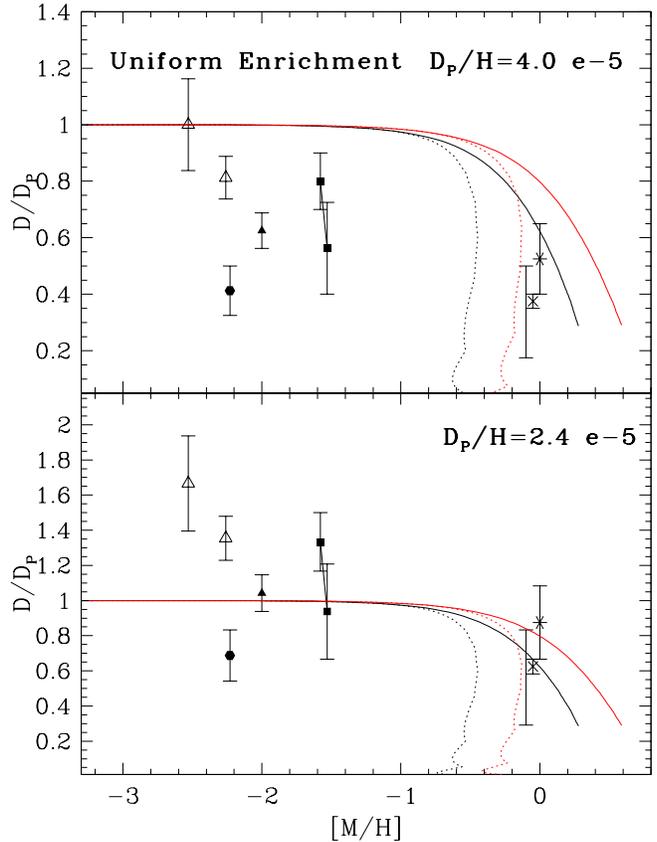,height=12.cm,width=0.5\textwidth}
\caption{\label{}
Deuterium vs. metallicity evolution, compared to observations of high
redshift regions, the protosolar nebula and the local ISM. Data in both
panels are: at high redshift from 
O'Meara et al. (2000, {\it triangles}), D'Odorico et al.
(2001, {\it square}), Levshakov et al. (2001, {\it square})
and Pettini \& Bowen (2001, {\it hexagon})
with {\it open symbols} corresponding to LLS and {\it filled symbols} to
DLAs; in the proto-solar nebula from Geiss (1998, {\it asterisk});
the local ISM from Linsky (1998, {\it cross}) and from recent IMAPS and FUSE
observations (Sonneborn et al. 2000, Vidal-Madjar 2001, 
range indicated by {\it vertical error bar}). 
In the {\it upper panel} data
is normalised to a primordial D value D$_P$/H = 4 10$^{-5}$ and in the {\it
lower panel}
to  D$_P$/H=2.4 10$^{-5}$ (see text).
Model curves in both panels are the same with
those in the middle panel of Fig. 1.
}
\end{figure}

\section {Discussion}

The number of D observations in high redshift systems is too small at
present to
allow for statistically significant conclusions about chemical evolution.
Still, it is interesting to see whether any inferences can be from the
available
data (taken at face value) about the evolutionary status of those systems
and/or
the primordial D abundance.

We note that the solutions of Eq. (1) and Fig. 1 concern the ratio \ddp \
and
not the absolute abundance D/H, whereas observations concern the latter
quantity.
Obviously, the higher the true primordial \dph \ value, the more difficult
is to deplete D down to the observed D/H values at high redshift.
This is illustrated  in the upper panel of Fig. 2, where all D/H values
(concerning high $z$ clouds, the proto-solar nebula and the local ISM)
have been normalised to \dph=4 10$^{-5}$. Although that value is still quite
modest (i.e. with respect to the high primordial values of \dph$>$10$^{-4}$
reported  by e.g . Webb et al. 1997), it is obvious that all ``standard''
models
systematically fail to reproduce the high redshift data, by more than 3
$\sigma$
in the case of the systems HS0105+1619 ({\it triangle}) and Q2206-197 ({\it
hexagone}).

In the lower panel of Fig. 2 all data are normalised to \dph=2.4 10$^{-5}$,
the weighted primordial D value found by averaging
over the available measurements of high column density systems  alone 
({\it filled symbols}). The data agree with the
theoretical curves in that case within 2 $\sigma$ in all cases. In our view,
this
is the most reasonable interpretation of the data at present, suggesting
that
DLAs have a undergone ``normal'' chemical evolution with negligible D
depletion
and reveal the true primordial D/H value (which may lay in the 2.-3
10$^{-5}$
range); as pointed out in Pettini \& Bowen (2001) the difficulty of
determining
the HI column density in LLS renders the D/H evaluation in those systems
prone
to systematic errors and thus less reliable.

\begin{figure}
\psfig{file=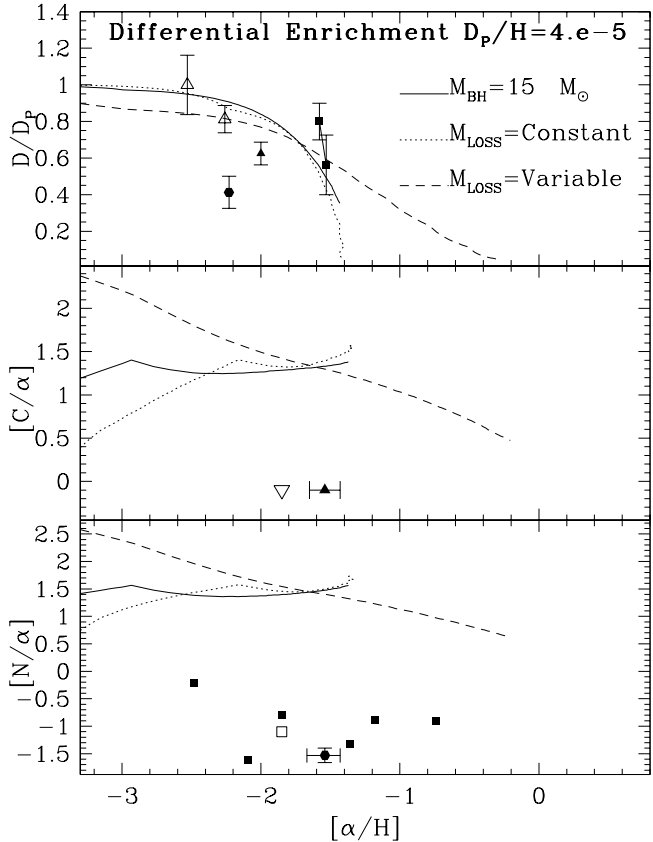,height=12.cm,width=0.5\textwidth}
\caption{\label{}
{\it Top:}
Deuterium vs. metallicity evolution, compared to observations of high
 redshift regions. Data are the same as Fig. 2, normalised to
D$_P$/H=4 10$^{-5}$. Curves correspond to ``non-standard'' models,
the same as in the bottom panel of Fig. 1. 
{\it Middle:}
[C/$\alpha$] vs. metallicity evolution. Curves correspond to the same models
as in the top panel; here and in the bottom panel
{\it filled symbols} represent abundance ratios where Si is the $\alpha$ element
while {\it open symbols} correspond to O.
Data point for $z$=4.46 DLA
({\it lower limit} for C/Si) is taken from Dessauges-Zavadsky et al. (2001)
and for HS 0105+1619 ({\it upper limit} for C/O) from O'Meara et al. (2001).
{\it Bottom:}
[N/$\alpha$] vs. metallicity evolution. Curves correspond to the same
models as in the top panel.
Data for DLAs are from Centurion et al. (1998,
{\it filled squares}), Dessauges-Zavadsky et al. (2001, {\it filled
hexagon}) and O'Meara et al. (2001, {\it open square}).
}
\end{figure}

We note that such low primordial D values are compatible with the pre-solar
value of D/H=2.5$\pm$0.5 10$^{-5}$ (Geiss 1998) and suggest that in the
solar neighborhood D has been virtually undepleted up to the Sun's
formation,
probably due to slow infall of primordial gas; such a slow infall is
necessary in order to explain the local G-dwarf metallicity distribution, as
argued
in several places (e.g. Prantzos \& Silk 1998, Tosi et al. 1998).

On the other hand, if one assumes that the slightly higher D/H values of LLS
reflect
the true primordial D abundance and that there is a trend of decreasing D
with
metallicity, then ``non-standard'' models have to be used to explain the
data, taken
at face value. This is seen in the top panel of Fig. 3, where it is assumed
that
\dph=4 10$^{-5}$. The data point of Pettini \& Bowen (2001) is difficult to
reproduce even with extreme assumptions about the system's history.
However, the common feature of {\it all models} obtaing substantial D
depletion
at low metallicities through astration
is that they have to minimize the amount of (metal-rich) ejecta
from massive stars while keeping the D-free ejecta of intermediate mass
stars
(either through ``selective winds'' or a skewed IMF).
As a result, typical nucleosynthetic products of  intermediate mass stars,
like C and N, should be particularly abundant in that case, as originally
suggested by Jedamzik \& Fuller (1997). 
This is illustrated in the middle and bottom panels of Fig. 3, where the
corresponding C/$\alpha$ and N/$\alpha$
ratios are found to be at least 10 times 
solar (where ``alpha'' stands for O or Si).
In our calculations we adopted the metallicity dependent yields of
Woosley \& Weaver (1995)
for massive stars and of Van den Hoek \& Groenewegen (1997) for
intermediate
mass  stars, which include Hot-Bottom Burning (HBB) and find substantial
nitrogen production in the 4-8 \ms \ stellar mass range.

The available data on C abundances in DLAs do not allow
at present a useful comparison to model predictions, since there is only one
lower limit
for the $z$=4.466 DLA towards the QSO APM BR J0307-4945 (Dessauges-Zavadsky
et al. 2001); since the column 
density of the cloud towards the QSO HS 0105+1619 is log(N$_{HI}$)=19.42, this
cloud could be marginally considered as a DLA, in which case the upper
limit on its  C/O ratio ({\it open symbol} in the middle panel of Fig. 3)
is clearly incompatible with the values of the ``non-standard'' senarii.
On the other hand, there are several observations of N/Si in DLAs (Centurion
et al. 1998,
Dessauges-Zavadsky et al. 2001) which show no particular enhancement of
nitrogen,
as can be seen in Fig. 3 (bottom panel). N/Si is considerably lower than
solar in
these systems, even if one takes into account ionisation corrections which
could enhance that ratio (e.g. Pilygin 1999 and references therein).
Taken at face value, the data do not favour the ``non-standard'' senario
of chemical evolution developed in this section.
One could argue, however, that HBB has not taken
place in intermediate mass stars of such low metallicity systems and that
nitrogen has
not to be necessarily enhanced. Indeed, the occurence (as a function
of stellar mass) and the amount of HBB are still
matters of debate (e.g. Lattanzio 1998 for a review). It turns out then that
carbon observations become crucial, since a large
C/Si (or C/O) ratio is always expected, independently of the occurence of HBB.

In summary, C/$\alpha$ {\it and/or} N/$\alpha$ (where $\alpha$ stands for
alpha-elements like O, Ne, Mg, Si, S etc.) 
should be particularly enhanced in high redshift
systems {\it if} their D has been indeed substantially depleted
through astration. The
only way to avoid the overproduction of {\it both} C/$\alpha$ and 
N/$\alpha$ is to assume that
in intermediate mass stars of such low metallicities the third dredge-up,
(bringing C from the He-burning shell to the stellar envelope), is supressed
(see Lattanzio 1998 and references therein).

The results of Fig. 3 also suggest that if the primordial D value is much
higher than
D$_P$/H $\sim$4 10$^{-5}$, then it is extremely difficult to intepret the
high redshift
data by any model invoking only astration in order to deplete D.
A possible loophole in that argument concerns the astration of D by
a first generation of superheavy (M$>$1000 \ms) stars, which would eject
in the ISM only their hydrogen-burning products (D free and He-rich) through
stellar winds before collapsing to black holes; this senario is advocated
in Jedamzik \& Fuller (1997) and cannot be refuted by current observations,
since the yields and the IMF of such stars are unknown. 
We note that this senario could accomodate both a high primordial D and a low
primordial He, since the currently determined values of those elements
in low metallicity systems would result from the action of those stars and
would not reflect directly the corresponding primordial values.
In that case the observational 
determination of primordial He abundance (through
the linear regression of He/H vs. O/H in low metallicity dwarf galaxies, e.g.
Izotov \& Thuan 1998) would become much less straightforward.

We also note that the apparent dispersion of D measurements in high $z$
absorbers (a factor of $\sim$2) is comparable to the one observed in the local
ISM (e.g. Vidal-Madjar 2001), which has not received a satisfactory 
explanation up to now. However, the reality of the dispersion at high $z$ is
not established yet, and even if it were the case, a different explanation 
would probably be required, in view of the vastly different spatial 
scales involved.

\section {Summary}

In this paper we argue that in any ``standard'' model of galactic chemical
evolution,
i.e. with a ``reasonable'' IMF covering the whole stellar mass range and
assuming
no preferential enrichment by low-mass stars, substantial D depletion
($>$20\%) can be
obtained only at metallicities [M/H]$>-1$; this conclusion is independent of
the
adopted SFR, infall or outflow prescriptions.

In the light of this argument, the recent observations of D/H vs.
metallicity
in several high redshift absorbers are best understood if the primordial D
value
is in the range D$_P$/H$\sim$2-3 10$^{-5}$, as suggested by O'Meara et al.
(2000) and
Pettini \& Bowen (2001) on the basis of D/H observations in DLAs. We
note that this range  points to a baryonic density
($\Omega_Bh^2$=0.019-0.026)
similar to that obtained by recent estimates based on the CMB anisotropy
measurements (Wang et al. 2001).

Slightly higher values
(D/H$\sim$4 10$^{-5}$) are found in Lyman limit systems. Such values
are barely compatible with CMB estimates and, if taken at face value,
suggest a trend of decreasing D abundance with metallicity. We argue that
special 
assumptions, like differential enrichment of DLAs, 
are required to explain the data
in that case. A clear test of such a differential enrichment would be
an excess of products of low mass stars like C and/or N in those systems
({\it if}
they have depleted D by astration alone, as assumed here). Currently
available data
of N/Si in DLAs do not favour such a ``non-standard'' senario, but
observations of
C/Si are crucial in order to definitively eliminate it.

\acknowledgements We are grateful to the referee, S. Burles, for his
constructive criticism. We acknowledge useful discussions with P. Petitjean.
Y. I. benefits from a post-doctoral fellowship of the french
Ministere de l'Education nationale, de la Recherche et de la Technologie
(Contract No: 353-2001).
\def\apj{ApJ}
\def\apjl{ApJL}
\def\apjs{ApJS}
\def\aj{AJ}
\def\aap{A\&A}
\def\araa{ARA\&A}
\def\aapss{A\&AS}
\def\mnras{MNRAS}
\def\nature{Nature}
\def\apss{Ap\&SS}
\def\pasp{PASP}

{}

\end{document}